\documentclass[aps,prb,showpacs,groupedaddress,floatfix,nofootinbib]{revtex4}
\usepackage{amsmath}

\usepackage{graphicx}
\usepackage{dcolumn}

\begin{document}

\title{Comment on: ``Development of the perturbation theory using polynomial
solutions'' [J. Math. Phys. \textbf{60}, 012103 (2019)]}

\author{Francisco M. Fern\'andez}\email{fernande@quimica.unlp.edu.ar}

\affiliation{INIFTA (CONICET, UNLP), Blvd. 113 y 64 S/N,
\\ Sucursal 4, Casilla de Correo 16, 1900 La Plata, Argentina}

\begin{abstract}
The purpose of this comment is to present the perturbation
approach proposed by Maiz [J. Math. Phys. \textbf{60}, 012103
(2019)] in a clearer way. The results of our straightforward
procedure agree with those obtained by that author except for one
case in which we obtain the exact result while he obtained an
approximate one. In addition to it, we show that for sufficiently
deep double-well potentials the perturbation approach deteriorates
considerably.
\end{abstract}

\pacs{03.65.Ge}

\maketitle

\section{Introduction}

\label{sec:intro}

In a recent paper Maiz\cite{M19} proposed a modified perturbation approach
in which the unperturbed or reference Hamiltonian operator is chosen to be
somewhat close to the actual, or perturbed, Hamiltonian operator. The
method, named exact polynomial potential solutions (EPPS from now on), was
restricted to one-dimensional polynomial potentials. Numerical results of
first order in perturbation theory appear to be reasonably accurate for a
family of anharmonic oscillators.

In our opinion the author presents the approach in a somewhat unclear and
confusing way. The purpose of this comment is to develop that perturbation
strategy more clearly. In section~\ref{sec:examples} we present the method
and apply it to the set of anharmonic oscillators discussed in that paper%
\cite{M19}. In section~\ref{sec:conclusions} we summarize the main
conclusions and show resuts for some examples not considered by Maiz.

\section{Exact polynomial potential solutions}

\label{sec:examples}

In what follows we just focus on the dimensionless Schr\"{o}dinger equation
\begin{eqnarray}
\psi ^{\prime \prime }(x) &=&\left[ V(x)-E\right] \psi (x),  \nonumber \\
V(x) &=&\sum_{i=1}^{N}b_{i}x^{i},  \label{eq:Schro}
\end{eqnarray}
where $N$ is an even number and $b_{N}>0$. Maiz\cite{M19} proposed an ansatz
of the form
\begin{eqnarray}
\psi (x) &=&f(x)\exp \left[ h(x)\right] ,  \nonumber \\
h(x) &=&\sum_{i=1}^{2N}a_{i}x^{i},  \label{eq:ansatz_Maiz}
\end{eqnarray}
where $f(x)=1$ for the ground state and $f(x)=\prod_{i=1}^{n}(x-x_{i})$ for
the excited states. In this way the author derived a quasi-exactly-solvable
Hamiltonian $H_{ex}$ and then improved the exactly known eigenvalue $E^{(0)}$
of $H_{ex}$ by means of perturbation theory of first order: $%
E^{(1)}=\left\langle H-H_{ex}\right\rangle $. It is clear that one needs a
recipe for the calculation of optimal values of the parameters in the
ansatz. As stated in the introduction we think that the author's
presentation of his method is rather unclear and the purpose of this comment
is to develop it in a somewhat clearer way.

The starting point is a square-integrable trial function $\varphi (x,\mathbf{%
a})$ where $\mathbf{a}$ is a set of adjustable parametes. From this function
we derive a potential $V_{0}(x,\mathbf{a})$ as
\begin{equation}
\frac{\varphi ^{\prime \prime }(x,\mathbf{a})}{\varphi (x,\mathbf{a})}%
=V_{0}(x,\mathbf{a})-E^{(0)}(\mathbf{a}).  \label{eq:V_0(x,a)}
\end{equation}
Then we obtain a correction of first order in the usual way
\begin{equation}
E^{(1)}=\frac{\int_{-\infty }^{\infty }\left[ V(x)-V_{0}(x,\mathbf{a}%
)\right] \varphi (x,\mathbf{a})^{2}dx}{\int_{-\infty }^{\infty }\varphi (x,%
\mathbf{a})^{2}dx}.  \label{eq:E^(1)}
\end{equation}
The accuracy of the result will obviously depend on the choice of the
adjustable parameters $\mathbf{a}$.

The equations developed above are quite general (in fact, once can easily
write similar equations for more than one dynamical coordinate), but in what
follows we restrict ourselves to the ground states of the polynomial
potentials considered by Maiz. To this end we choose
\begin{eqnarray}
\varphi (x,\mathbf{a}) &=&\exp \left[ h(x,\mathbf{a})\right] ,  \nonumber \\
h(x,\mathbf{a}) &=&\sum_{i=1}^{M}a_{i}x^{i},  \label{eq:ansatz}
\end{eqnarray}
where $M$ is even and $a_{M}>0$. The potential of order zero
\begin{equation}
V_{0}(x,\mathbf{a})=h^{\prime }(x,\mathbf{a})^{2}-h^{\prime \prime }(x,%
\mathbf{a})+E^{(0)}(\mathbf{a}),  \label{eq:V_0(x,a)_b}
\end{equation}
where $E^{(0)}(\mathbf{a})=2a_{2}-a_{1}^{2}$ and the prime stands for
derivative with respect to $x$, is a polynomial function of order $2M-2$. We
arbitrarily choose the $M$ adjustable parameters $\mathbf{a}%
=(a_{1},a_{2},\ldots ,a_{M})$ in order to remove $M$ terms of the
perturbation potential $V(x)-V_{0}(x,\mathbf{a})$. In order to reproduce the
results of Maiz we choose $2M-2>N$ and, obviously, $a_{2j+1}=0$ if $%
V(-x)=V(x)$.

Our first example is the exactly solvable harmonic oscillator $V(x)=x^{2}$.
If $M=4$ we have
\begin{equation}
V(x)-V_{0}(x,\mathbf{a})=-16a_{4}^{2}x^{6}-16a_{2}a_{4}x^{4}-x^{2}\left(
4a_{2}^{2}-12a_{4}-1\right) ,  \label{eq:[V-V_0]_HO}
\end{equation}
from which it follows that $a_{4}=0$ and $a_{2}=1/2$. We thus obtain, as
expected, the exact ground state energy at zero order $E^{(0)}=1$. Curiously
enough, Maiz\cite{M19} obtained an approximate result for this trivial case.
At first sight it may seem that both approaches are different.

In the case of $V(x)=x^{4}$ we choose the same ansatz and obtain the
following results
\begin{eqnarray}
a_{2} &=&\frac{12^{1/3}}{4},\;a_{4}=\frac{18^{1/3}}{24},  \nonumber \\
V(x)-V_{0}(x,\mathbf{a}) &=&-\frac{12^{1/3}}{12}x^{6}=-0.190785707x^{6},
\nonumber \\
E_{0} &=&\frac{12^{1/3}}{2}=1.144714242,  \label{eq:x^4}
\end{eqnarray}
that exactly agree with those of Maiz\cite{M19}. The calculation of $%
E^{(1)}=-0.07198347757$ should be carried out numerically and we appreciate
that it also agrees with the result of that author. We thus have $%
E=E^{(0)}+E^{(1)}=1.072730764$ that closely agrees with accurate results
obtained by other means\cite{M19} (see also Table~\ref{tab:reference} to be
discussed later on). Now it seems that our approach is identical to EPPS but
we have not been able to find the source of the discrepancy in the case of
the harmonic oscillator.

In the case of the potential $V(x)=x^{2}+x^{3}+x^{4}$ we also choose $M=4$
but we include both even- and odd-parity terms. A straightforward numerical
calculation shows that
\begin{eqnarray}
a_{1}
&=&0.22892176,\;a_{2}=0.6805239186,\;a_{3}=0.1038578221,%
\;a_{4}=0.08292525897,  \nonumber \\
E_{0} &=&1.308642664,\;V(x)-V_{0}(x,\mathbf{a}%
)=-0.206698483x^{5}-0.1100255772x^{6},  \label{eq:x2x3x4}
\end{eqnarray}
in agreement with the results of EPPS, except for a slight discrepancy in
the coefficient of $x^{6}$ that is probably due to numerical errors. We also
obtain $E^{(1)}=0.004710353228$ and $E=1.313353017$ in perfect agreement
with EPPS.

Present approach agrees with EPPS also in the case of the other
models considered in that paper\cite{M19}. In general, the results
provided by this approach appear to be quite reasonable. In order
to test the accuracy of his results Maiz resorted to reference
eigenvalues obtained by means of other approach\cite{M14}.
However, in that paper there are results only for even-parity
potentials. Table~\ref{tab:reference} shows accurate benchmark
energies calculated by means of the Riccati-Pad\'{e} method
(RPM)\cite {FMT89b,FT96}. We appreciate that Maiz's reference
eigenvalues are less accurate than the reported number of
significant digits appears to suggest.

\section{Further comments and conclusions}

\label{sec:conclusions}

It has been our purpose in this comment to develop the EPPS in a simpler and
clearer way. In doing so we found out that Maiz\cite{M19} should have
obtained the exact result for the harmonic oscillator instead of the
approximate one reported in his paper. The EPPS appears to yield reasonable
results for the ground states of polynomial potentials by means of
first-order perturbation theory. However, for such simple models one easily
obtains very accurate results by means of the straighforward Raleigh-Ritz
variational method or any other approach. Here we have resorted to the RPM%
\cite{FMT89b,FT96} that converges exponentially fast. The results in Table~%
\ref{tab:reference} were otained from the roots of Hankel determinants of
dimension $D\leq 10$.

In the case of double-well potentials the EPPS may perform poorly if the
wells are sufficiently deep. For example, in the case of $V(x)=x^{4}-\lambda
x^{2}$ we obtain ($E^{[EPPS]}=0.7122694296,$ $E^{[RPM]}=0.65765300518071512$%
) for $\lambda =1$ and ($E^{[EPPS]}=0.885893999,$ $E^{[RPM]}=0.63891956378$)
for $\lambda =5$. The percent errors $8.3$ and $38.7$, respectively, are
considerably larger than those for the models chosen by Maiz\cite{M19}.

\begin{table}[tbp]
\caption{Reference ground-state eigenvalues calculated by means of two
different approaches}
\label{tab:reference}
\begin{center}
\par
\begin{tabular}{lD{.}{.}{8}D{.}{.}{15}}
\hline \multicolumn{1}{c}{$V(x)$}&
\multicolumn{1}{c}{Maiz\cite{M14}} &
\multicolumn{1}{c}{RPM}  \\
\hline

$x^4$                     &  1.06065    &   1.0603620904841829 \\
$x^2+x^3+x^4$           &  1.310342   &    1.31025752970575  \\
$x^6$                   &  1.14571    &   1.14480245380      \\
$x^2+x^6$               &  1.43555    &   1.43562461900      \\
$x^4+x^5+x^6$           &  1.3032     &  1.30272754246       \\
$x^2-x^3+x^4+x^6$       &  1.586428   &    1.58657805318     \\
$x^2-x^3+x^4-x^5+x^6$   &  1.470961   &    1.4711571858      \\
\hline
\end{tabular}
\par
\end{center}
\end{table}

\end{document}